\documentclass[epj]{svjour}
\usepackage{graphics}
\begin{document}
\title{Isospin breaking in the vector current of the nucleon}
\author{Randy Lewis
}                     
%
%
\institute{Department of Physics, University of Regina, Regina, Saskatchewan,
           Canada, S4S 0A2}
%
\date{31 August 2006}
%
\abstract{
Extraction of the nucleon's strange form factors from experimental data
requires a quantitative understanding of the unavoidable contamination from
isospin violation.
A number of authors have addressed this issue during the past decade, and
their work is reviewed here.
The predictions from early models are largely consistent with recent results
that rely as much as possible on input from QCD symmetries and related
experimental data.
The resulting bounds on isospin violation are sufficiently precise to be of
value to on-going experimental and theoretical studies of the nucleon's strange
form factors.
} 
\maketitle
\section{Motivation} \label{sec:mot}

Isospin violation is generally a small effect.  For example,
consider the nucleon mass splitting, $(m_n-m_p)/m_p=0.1\%$.
One similarly expects isospin violation to have a small impact
on the nucleon's electromagnetic and weak form factors.
However, this does not imply
that iso\-spin violation must be small relative to strangeness effects.
To illustrate, recall that an explicit calculation in the electroweak
theory leads to 
\begin{eqnarray}
G_X^{p,Z}(q^2) = \hspace{67mm} \nonumber \\
 (1-4\sin^2\theta_W)G_X^p(q^2) - G_X^n(q^2) - G_X^s(q^2) - G_X^{u,d}(q^2)
\hspace{3mm} \label{maindef}
\end{eqnarray}
for electric ($X=E$) and magnetic ($X=M$) form factors.  Experimental
studies\cite{SAMPLE,A4a,A4b,HAPPEXa,HAPPEXb,G0} show that the sum of the
last two terms on the right-hand
side is small.  The size of isospin violation, $G_X^{u,d}(q^2)$, relative to
strangeness, $G_X^s(q^2)$, is not obtained from these experiments.

In what follows, theoretical studies of $G_X^{u,d}(q^2)$ will be
reviewed\cite{DP,Mi,Ma,LM,KL}. (Our entire discussion of isospin
violation also fits within the more restrictive category called ``charge
symmetry breaking'' and that language is used, for example, in Ref.~\cite{Mi}.)
If the current understanding of these isospin violating effects is sufficiently
precise, then the data from
Refs.~\cite{SAMPLE,A4a,A4b,HAPPEXa,HAPPEXb,G0}
allow for a determination of the authentic
strange quark effects, $G_X^s(q^2)$, which are of great interest to many people
at present.

Independent of any chosen theoretical approach,
each isospin violating form factor is simply
the difference of isoscalar and isovector terms,
\begin{equation}
G_X^{u,d}(q^2) \equiv G_X^{s\!\!\!/}(q^2) - G_X^{v\!\!\!/}(q^2)\,,
\end{equation}
where $G_X^{s\!\!\!/}$ is obtained from
\begin{equation}\label{isoscalar}
\left<p\left|\bar{u}\gamma_\mu u+\bar{d}\gamma_\mu d\right|p\right>
-
\left<n\left|\bar{u}\gamma_\mu u+\bar{d}\gamma_\mu d\right|n\right>
\end{equation}
and $G_X^{v\!\!\!/}$ is obtained from
\begin{equation}\label{isovector}
\left<p\left|\bar{u}\gamma_\mu u-\bar{d}\gamma_\mu d\right|p\right>
+
\left<n\left|\bar{u}\gamma_\mu u-\bar{d}\gamma_\mu d\right|n\right>
\end{equation}
in a straightforward manner (see Refs.~\cite{DP,KL} for details).
Furthermore, we know that all isospin violation is ultimately a consequence of
unequal quark masses, $m_u\neq m_d$, (``strong breaking'') and unequal quark
electric charges, $e_u\neq e_d$ (``electromagnetic breaking'').
The task for each theoretical approach is to determine the
combinations of nucleon matrix elements shown in Eqs.~(\ref{isoscalar}) and
(\ref{isovector}), with both types of breaking included.

Since the sum of strangeness and isospin violation in Eq.~(\ref{maindef}) is
measured to be a small fraction of the total form factors, and since isospin
violation itself is expected to be a small fraction of the total form factors,
it is reasonable to neglect contributions containing {\em both}
strangeness and isospin violation as doubly (i.e. negligibly) small.
This allows $G^{u,d}_X(q^2)$ to be calculated without dynamical strange-quark
effects.  Such an approach is clearly advantageous for chiral
perturbation theory, where addition of a dynamical strange quark leads to
severe degradation of convergence properties of the chiral expansion.
All of the theoretical studies to date have computed isospin violation without
dynamical strange quarks.

Our discussion will be approximately chronological.
The constituent quark model studies of Dmitra\v{s}inovi\'c and
Pollock\cite{DP} and Mill\-er\cite{Mi},
are discussed in Sec.~\ref{sec:qm}, followed by Ma's use of a light-cone
meson-baryon fluctuation
mod\-el\cite{Ma} in Sec.~\ref{sec:lcm}.  The constraints of chiral symmetry
are discussed in Sec.~\ref{sec:chpt}, based on a collaboration with
Mobed\cite{LM} using chiral perturbation theory.
Section~\ref{sec:res} reviews the recent results from work with Kubis\cite{KL}
that combines
chiral perturbation theory
with resonance saturation and information from dispersion analyses.
The final section, Sec.~\ref{sec:sum}, provides a brief summary.

\section{Constituent quark model} \label{sec:qm}

In a constituent quark model, we might expect the scale of strong breaking
to be set by {\em constituent} quark masses, which the authors of
Ref.~\cite{DP} take to be $(m_D-m_U)/m_Q\sim1.2\%$, and
the scale of electromagnetic breaking to be set by $\alpha\sim0.7\%$.
To do better than this order-of-magnitude guess, an explicit calculation is
required.

The first calculation was carried out by Dmitra\v{s}inovi\'c and
Pollock\cite{DP} using oscillator confinement and a Coulomb potential,
\begin{eqnarray}
H &=& H_0 + V_{EM}\,, \\
H_0 &=& \sum_{i=1}^3\frac{{\bf p}_i^2}{2m_i}
  + \frac{k}{2}\sum_{i<j}^3({\bf r}_i-{\bf r}_j)^2\,, \\
V_{EM} &=& \sum_{i<j}^3\frac{e_ie_j}{4\pi\left|{\bf r}_i-{\bf r}_j\right|}\,.
\end{eqnarray}
Recall that this choice for $H_0$ produces Gaussian spatial wave functions.
With parameters fixed to be $m_Q=330$ MeV, $m_D-m_U=4$ MeV, and $k$ determined
from the experimental mass difference between $\Delta(1232)$ and nucleon,
Ref.~\cite{DP} finds
\begin{eqnarray}
G_M^{s\!\!\!/}(0) = G_M^{v\!\!\!/}(0)
&=& \frac{1}{3}\left[\frac{m_D}{m_U}-\frac{m_U}{m_D}\right] \approx 0.008\,, \\
G_M^{u,d}(0) &=& G_M^{s\!\!\!/}(0) - G_M^{v\!\!\!/}(0) = 0\,, \label{DP0} \\
\frac{\delta\left<r_E^2\right>}{\left<r_E^2\right>} &=& 1.1\%\,, \\
\frac{\delta\left<r_M^2\right>}{\left<r_M^2\right>} &=& 0.4\%\,,
\end{eqnarray}
where $r_X^2$ denotes a squared radius as usual.
These estimates are compatible with our order-of-magnitude guesses, but
what confidence level should be assigned to the precise values?

Here is a list of some limitations of this model
(all of which are mentioned explicitly in Ref.~\cite{DP}):
\begin{itemize}
\item The chosen parameters lead to, $m_n-m_p=3$ MeV, 230\% above experiment.
\item The nucleon charge radius,
 $\sqrt{\left<r_E^2\right>}=0.62$ fm, is 30\% below experiment.
\item Gaussian spatial wave functions cause both $G_E(q^2)$ and $G_M(q^2)$
      to be unrealistic at large ${\bf q}^2$.
\item The strong hyperfine interaction is omitted.
\item Chiral symmetry is absent.
\end{itemize}
In Section IV B of Ref.~\cite{DP}, the authors conclude,
``\ldots we may expect to have calculated
the correct sign and order of magnitude of the effects of interest.''

Subsequently, Miller\cite{Mi} used a more complete constit\-u\-ent quark model
and explicitly addressed each of the limitations listed above, except
chiral symmetry.
(Miller says chiral symmetry is implicit in
the charge symmetry conserving pion cloud of
this model\cite{private}.)
His Hamiltonian is
\begin{eqnarray}
H &=& K + V_{\rm con} + V_{\rm em} + V_g\,, \hspace{4cm} \\
K &=& \sum_{i=1}^3\left(m_i+\frac{p_i^2}{2m_i}\right)\,,
\end{eqnarray}
\begin{equation}
V_{\rm em} = \alpha\sum_{i<j}q_iq_j\left(\frac{1}{r_{ij}}-\frac{\pi}{2}
             \delta(\vec r_{ij})\left[\frac{2}{\bar{m}^2}+\frac{4}{3}
             \frac{\vec\sigma(i)\cdot\vec\sigma(j)}{\bar{m}^2}\right]\right)\,,
\end{equation}
\begin{equation}
V_g = -\alpha_s\!\!\sum_{i<j}\lambda_i\cdot\lambda_j\!\!\left[\frac{\pi}{2}
        \delta(\vec r_{ij})\!\left(\frac{1}{m_i^2}+\frac{1}{m_j^2}+\frac{4}{3}
        \frac{\vec\sigma(i)\cdot\vec\sigma(j)}{m_im_j}\right)\right]\!.
\end{equation}
Two options for $V_{\rm con}$ are studied; one gives Gaussian form factors,
\begin{eqnarray}
\Psi(\rho,\lambda) &=& N\exp{\left(\frac{\rho^2+\lambda^2}{-2\beta}\right)}\,,
 \\
\vec\rho &=& \frac{1}{\sqrt{2}}(\vec r_1-\vec r_2)\,, \\
\vec\lambda &=& \frac{1}{\sqrt{6}}(\vec r_1+\vec r_2-2\vec r_3)\,,
\end{eqnarray}
and the other gives power-law form factors,
\begin{eqnarray}
&& \Psi^2(R) = \frac{2\sqrt{6}\Lambda^7}{\pi^36^5}RK_1\left(\sqrt{\frac{2}{3}}
            \Lambda R\right)\,, \nonumber \\
\Rightarrow &&
G_E(Q^2) = \left(\frac{\Lambda^2}{Q^2+\Lambda^2}\right)^4\,.
\end{eqnarray}

For the Gaussian case, Ref.~\cite{Mi} uses the experimental proton magnetic
moment to determine $\bar m=337$ MeV, the experimental $m_\Delta-m_N$ value
to fix $\alpha_s$ as a function of $\beta$, and then considers three choices
for $\beta$ which define three models as shown in Table~\ref{tab:Mi}.
\begin{table}
\caption{Parameter values for the three Gaussian models of Ref.~\cite{Mi}.}
\label{tab:Mi}
\begin{center}
\begin{tabular}{llll}
\hline\noalign{\smallskip}
Model & 1 & 2 & 3 \\
\noalign{\smallskip}\hline\noalign{\smallskip}
$\sqrt{\beta}$ (fm) & 0.7 & 0.6 & 0.5 \\
$\alpha_s$ & 2.3 & 1.2 & 0.35 \\
$m_D-m_U$ (MeV) & 5.2 & 3.8 & 2.3 \\
\noalign{\smallskip}\hline
\end{tabular}
\end{center}
\vspace{-3mm}
\end{table}
Figures 4, 5, 6 and 7 of Ref.~\cite{Mi} display the resulting effects of
isospin violation on $G_E(q^2)$ and $G_M(q^2)$, including the separate
contributions from $K$, $V_{\rm em}$ and $V_g$.  In each case, the isospin
violating contributions vanish for $q^2=0$, and are less than 0.2\% in
magnitude for a momentum transfer of 0.1 GeV$^2$.
For power-law form factors, Ref.~\cite{Mi} obtains $\Lambda=5.90$/fm from the
experimental value of $\left<r_E^2\right>$, and all isospin violating effects
remain small.

The combined work of Refs.~\cite{DP,Mi} provides an excellent
understanding of isospin violation
within the constituent quark model, but we must now ask which features of the
results are a true reflection of nature, and which are model-dependent.
For example, there are symmetries in these quark models that lead to
$G_M^{u,d}(0)=0$, recall Eq.~(\ref{DP0}), but, as we'll see in
Sec.~\ref{sec:chpt}, this is not a
symmetry of nature.  It is therefore interesting to explore other theoretical
approaches as well.

\section{Light-cone meson-baryon fluctuation model} \label{sec:lcm}

To discuss the method used by Ma\cite{Ma}, we must think
at the hadron level rather than the constituent quark level.
Fluctuations of a nucleon into a virtual pion-nucleon pair,
$p\to\pi^+n$ and $n\to\pi^-p$, occur commonly.
As always, isospin violation arises from strong breaking (at the hadron level,
this means $m_n-m_p\neq0$) and from electromagnetic breaking (which now means
the Coulomb attraction of a charged/charged $\pi^-p$ pair is different from the
charged/neutral $\pi^+n$ pair).
The expression for isospin violation in the magnetic form factor at vanishing
momentum transfer, is
\begin{equation}\label{Ma0}
G_M^{u,d}(0) = \bigg(P(n\to\pi^-p)-P(p\to\pi^+n)\bigg)
               \bigg(\mu^n_{\pi^-p}-\mu_n\bigg)\,,
\end{equation}
where $P()$ denotes a fluctuation probability, and $\mu^n_{\pi^-p}$ is the
magnetic moment for the neutron's fluctuation.

To determine $\mu_{\pi^-p}^n$, begin with the fact that
total angular momentum is orbital angular momentum plus proton spin,
\begin{equation}
\left|\frac{1}{2},\frac{1}{2}\right>_J
= \sqrt{\frac{2}{3}}\bigg|1,1\bigg>_L
  \left|\frac{1}{2},-\frac{1}{2}\right>_S
- \sqrt{\frac{1}{3}}\bigg|1,0\bigg>_L
  \left|\frac{1}{2},\frac{1}{2}\right>_S\,.
\end{equation}
This leads directly to
\begin{eqnarray}
\mu_{\pi^-p}^n &=& -\frac{\mu_p}{3}
             + \frac{\mu_n}{3}\left(\frac{2m_n}{m_{\pi^-}+m_p}\right)
               \left(\frac{m_{\pi^-}}{m_p}-\frac{m_p}{m_{\pi^-}}\right)
               \nonumber \\
             &=& -4.75~~~~~~
\Rightarrow~~~~~~\mu_{\pi^-p}^n - \mu_n = -2.84\,.
\end{eqnarray}

To determine the difference of fluctuation probabilities needed for
Eq.~(\ref{Ma0}), a light-cone Gaussian wave function is used,
\begin{equation}
\psi = Ae^{-({\cal M}^2-m_N^2)/(8\alpha^2)}\,,
\end{equation}
where
\begin{equation}
{\cal M}^2=\sum_{i=1}^2\frac{{\bf k}_{\perp\,i}^2+m_i^2}{x_i}
\end{equation}
is the invariant mass of the meson-baryon state.
(Note the implicit assumption that $A$ is independent of which nucleon is
fluctuating; Ref.~\cite{Ma} points out that this assumption is not required,
but the uncertainties associated with relaxing it are difficult to estimate.)
{}From the experimental Gottfried sum rule, one finds
\begin{equation}
P(p\to\pi^+n)\approx P(n\to\pi^-p)\approx0.15\,.
\end{equation}

The final remaining parameter is the radius, $\alpha$, and it leads to
a large uncertainty.
Ref.~\cite{Ma} uses two bounds,
\begin{equation}
\alpha=300 {\rm MeV} ~~~\Rightarrow~~~ P(n=\pi^-n)-P(p=\pi^+n)=0.2\%\,,
\end{equation}
and
\begin{eqnarray}
\left.\begin{array}{l} \alpha(n\to\pi^-p)=205 {\rm MeV} \\
                       \alpha(p\to\pi^+n)=200 {\rm MeV} \end{array}
\right\} && \nonumber \\
    && \hspace{-2cm} \Rightarrow~~~ P(n=\pi^-n)-P(p=\pi^+n)=3\%\,, \nonumber \\
\end{eqnarray}
to arrive at
\begin{equation}
G_M^{u,d}(0) \approx 0.006\to0.088\,.
\end{equation}
However, Miller suggests that (205-200)/200=2.5\% is too large for a Coulomb
effect, and states: ``A reasonable estimate of the effect could be 0.03 nuclear
magnetons.''\cite{Mi}.

Thus, the light-cone meson-baryon fluctuation model provides an explicit
example of a model that does not lead to $G_M^{u,d}(0)=0$.  In particular,
this model predicts $G_M^{u,d}(0)>0$.

\section{Chiral perturbation theory} \label{sec:chpt}

Chiral perturbation theory is not a model; it merely administrates the global
symmetries of QCD.  This has the advantage of retaining no model dependence
in predictions, and it has the disadvantage that quantities not determined
by global symmetries remain completely unknown.  In Ref.~\cite{LM}, isospin
violation in the nucleon's vector form factors is studied within chiral
perturbation theory to leading order for the electric case, and to
next-to-leading order for the magnetic case.

Because isospin violation arises from both strong breaking and electromagnetic
breaking, our chiral perturbation theory must include dynamical photons as well
as dynamical pions.  A single baryon will flow through the process, though its
identity may change from proton to neutron or even to
$\Delta(1232)$ by absorbing
or emitting a pion (recall the fluctuations discussed in Sec.~\ref{sec:lcm}).
Ref.~\cite{LM} uses the formalism of heavy baryon chiral perturbation
theory\cite{JM,BKM}, and the perturbative expansion is defined in powers of 
electric charge $e$, momentum $q/\Lambda$, pion mass $m_\pi/\Lambda$, and
mass difference $(m_\Delta-m_N)/\Lambda$, where $\Lambda$ denotes either
$4\pi F_\pi$ or $m_N$.  Writing the Lagrangian in standard notation,
\begin{equation}
{\cal L}_{\rm ChPT} = {\cal L}^{(1)} + {\cal L}^{(2)} + {\cal L}^{(3)}
                    + {\cal L}^{(4)} + {\cal L}^{(5)} + \ldots\,,
\end{equation}
Ref.~\cite{LM} shows that the isospin violating form factors begin in
${\cal L}^{(4)}$ (plus the corresponding loop diagrams), and
next-to-leading effects are in ${\cal L}^{(5)}$ (plus the corresponding
loop diagrams).
For a general observable at this high order, two-loop diagrams routinely appear
and the most general effective Lagrangian contains literally
hundreds of low-energy constants (i.e. parameters whose numerical values
are not constrained by global symmetries, and hence unknown to chiral
perturbation theory).

For the specific case of $G_M^{u,d}(q^2)$, Ref.~\cite{LM} shows that the
situation is much simpler than for a general observable due to five key
observations:
\begin{itemize}
\item $G_M^{u,d}(q^2)$ receives no two-loop contributions up to next-to-leading
      order.  (In particular, ``photon+photon'' loop diagrams lack the required
      spin operators, ``photon+ pion'' loop diagrams sum to
      {\em anticommutators} of
      Pauli-Lubanski operators, and ``pion+pion'' loop diagrams have only
      higher-order isospin violation.)
\item All one-loop photon effects in $G_M^{u,d}(q^2)$ can be absorbed into
      the physical value of $m_n-m_p$.
\item Explicit $\Delta(1232)$ effects are found, numerically, to be smaller
than systematic uncertainties.
\item $G_M^{u,d}(q^2)$ contains just one combination of low-energy constants
      at leading order --- it's a simple additive constant --- and no
      additional low-energy constants at next-to-leading order.
      (The fact that no global symmetry forces this constant to vanish is not
      realized in some constituent quark models, as discussed in
      Sec.~\ref{sec:qm}.)
\item All isospin breaking in loops, up to next-to-leading order, is ultimately
      due to $m_n-m_p$.
\end{itemize}
Based on these observations, a parameter-free prediction for the difference
$G_M^{u,d}(q^2)-G_M^{u,d}(0)$ is obtained in Ref.~\cite{LM}.

Isospin violation in the electric form factor is discussed in
Ref.~\cite{MENU99}, but only to leading
order because in this case there {\em are} two-loop contributions and
additional unknown constants at next-to-leading order.

Though it is useful to have these ``pion cloud contributions'' computed
within chiral perturbation theory, the missing combination of low-energy
constants indicates that chiral perturbation theory alone will not meet all
of our goals, and we must turn our attention to a method for understanding the
physics that resides within the low-energy constants.

\section{Chiral perturbation theory with resonance saturation} \label{sec:res}

The chiral perturbation theory calculations of Ref.~\cite{LM} were
reproduced in Ref.~\cite{KL} using two separate formalisms: a repeat of the
heavy baryon calculation, and a newer method known as infrared
regularization\cite{BL}.  This newer approach is simpler to
manage and needs fewer Feynman diagrams, but physical results must be
identical.  Ref.~\cite{KL} finds and corrects some errant factors of 2 in
Ref.~\cite{LM}, and confirms all of the features mentioned in
Sec.~\ref{sec:chpt}.

\begin{figure}
\begin{center}
\resizebox{0.45\textwidth}{!}{\includegraphics{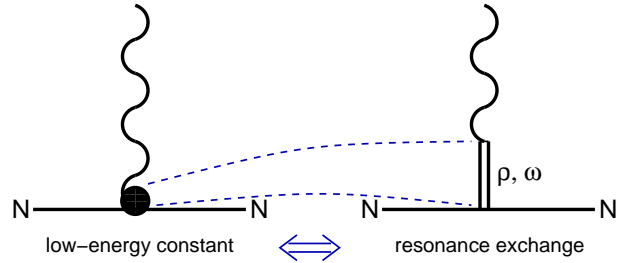}}
\end{center}
\caption{The basic idea of resonance saturation for a nucleon's vector form
         factor.}\label{fig:ressat}
\end{figure}

With the results of chiral perturbation theory now firmly in hand,
Ref.~\cite{KL} addresses the issue of unknown low-energy constants.
They are not constrained by chiral symmetry,
nor by any other global symmetry of QCD.
Since chiral perturbation theory is the low-energy effective theory of QCD,
where pions (and in our case, photons too) are dynamical, the low-energy
constants must account for all of the higher-energy QCD dynamics.  Most
important would be the exchange of the lightest resonances that are too
heavy to appear explicitly in the effective theory: for vector form factors,
these are the $\rho$ and $\omega$.  Figure \ref{fig:ressat} sketches this
notion of resonance saturation.
Mathematically, it is easy to see that the exchange of a heavy particle is
a low-energy constant at leading order,
\begin{equation}
{\rm propagator} \sim
\frac{1}{M^2-q^2}=\frac{1}{M^2}+O\left(\frac{q^2}{M^2}\right)\,.
\end{equation}
Of course these resonances have to be coupled in a way consistent with
chiral symmetry.

This type of resonance saturation was shown long ago to work very well in the
meson chiral Lagrangian\cite{ressat1,ressat2,ressat3}, though it has been shown
that the nucleon's vector form factors require inclusion of extra resonances
beyond merely the
$\rho$ and $\omega$\cite{KM}.  Nevertheless, our present goal is
only to compute the isospin violating pieces of these form factors; they
come from $\rho-\omega$ mixing, and Ref.~\cite{KL} points out that any
effects of higher resonances are more severely suppressed in this case.

\begin{figure}
\begin{center}
\resizebox{0.45\textwidth}{!}{\includegraphics{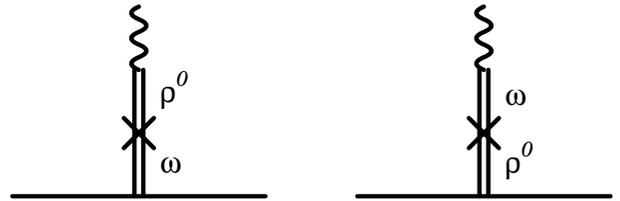}}
\end{center}
\caption{Resonance saturation contributions to isospin violation in the
         nucleon's vector form factors.}\label{fig:mixing}
\end{figure}

The contributions of resonance saturation to the isospin violating form factors
are shown diagrammatically in Fig.~\ref{fig:mixing}.
The corresponding equations are
\begin{eqnarray}
\delta G_E^{u,d}(q^2) &=& 
   \frac{\Theta_{\rho\omega}}{M_V(M_V^2-q^2)^2}
   \left[
   \left(1 + \frac{\kappa_\omega M_V^2}{4m_N^2}\right)g_\omega F_\rho q^2
   \right. \nonumber \\
&& \left.
 - \left(1 + \frac{\kappa_\rho M_V^2}{4m_N^2}\right)g_\rho F_\omega q^2
   \right]\,,
   \label{GEmixing} \\
\delta G_M^{u,d}(q^2) &=&
     \frac{\Theta_{\rho\omega}}{M_V(M_V^2-q^2)^2}
     \left[
     \left(q^2 + \kappa_\omega M_V^2\right)g_\omega F_\rho
     \right. \nonumber \\
 &&  \left. 
   - \left(q^2 + \kappa_\rho M_V^2\right)g_\rho F_\omega
     \right]\,,
     \label{GMmixing}
\end{eqnarray}
where $M_V$, $F_\rho$ and $F_\omega$ are the vector meson mass and decay
constants, and the $\rho-\omega$ mixing parameter has been determined from
experimental masses and branching ratios to be\cite{KucM}
\begin{equation}
\Theta_{\rho\omega}=(-3.75\pm0.36)\times10^{-3}{\rm GeV}^{-2}\,.
\end{equation}

If we had numerical values for the couplings of vector mesons to nucleons,
$g_\rho$, $\kappa_\rho$, $g_\omega$ and $\kappa_\omega$, then
Eq.~(\ref{GMmixing}) could be expanded in powers of $q^2$ to obtain
the desired low-energy constant (at $q^2=0$) as well as additional
contributions that are technically of higher order in the chiral expansion.
We immediately see that Eq.~(\ref{GEmixing}) vanishes at $q^2=0$ as required
by global symmetries, but nonzero contributions from higher orders in the
chiral expansion would be obtained.
\begin{figure}
\vspace{4mm}
\begin{center}
\resizebox{0.45\textwidth}{!}{\includegraphics{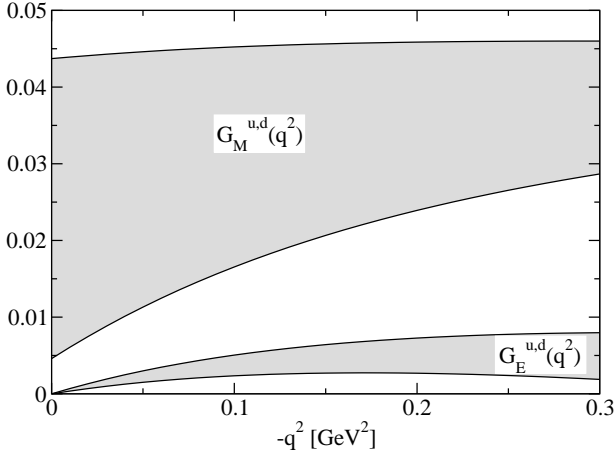}}
\end{center}
\caption{Isospin violation in the electromagnetic form factors, as obtained
         from Ref.~\protect\cite{KL}.}\label{fig:final}
\end{figure}

In principle, there is a wide assortment of techniques for attempting
to quantify $g_\rho$, $\kappa_\rho$, $g_\omega$ and $\kappa_\omega$, but
finding an approach with rigorously-quantifiable uncertainties is difficult.
In an attempt to rely as directly as possible on experimental data rather
than models, Ref.~\cite{KL} uses values extracted from
dispersive analyses of nucleon electromagnetic form factors.
For the $\rho$ couplings, these analyses must account for the non-resonant
two-pion continuum in addition to the Breit-Wigner $\rho$ resonance.  Data from
Refs.~\cite{MMD,BHM} lead to the ranges
\begin{eqnarray}
4.0< & g_\rho& <6.2\,, \label{grho} \\
5.1< & \kappa_\rho& <6.8\,.
\end{eqnarray}
For the $\omega$ couplings, it is sufficient to consider only pure zero-width
resonance pole residues, and Refs.~\cite{MMD,HM,BHM2} produce the ranges
\begin{eqnarray}
41.8< & g_\omega& <43.0\,, \\
-0.16< & \kappa_\omega& <0.57\,. \label{komega}
\end{eqnarray}
The final results of Ref.~\cite{KL}, reproduced in Fig.~\ref{fig:final},
include the worst-case errors bars obtained from spanning the ranges in
Eqs.~(\ref{grho}-\ref{komega}) above; the poorly-known $\kappa_\omega$
dominates the uncertainties.  Notice that the range of $G_M^{u,d}(0)$ does not
include zero, while $G_E^{u,d}(0)=0$ is required.  Both form factors are
positive over the momentum range considered.
\vspace{-3mm}

\section{Summary} \label{sec:sum}

Early studies of isospin violation in the nucleon's vector form factors
led to a clear understanding within specific quark models.
Use of chiral perturbation theory avoids all model dependence, but leaves some
parameters undetermined.  Phenomenologically, those parameters are saturated
by resonances, and numerical values are obtained
with minimal model-dependence
by using dispersive analyses.
The results in Fig.~\ref{fig:final} represent a conservative determination
of isospin violation, as obtained from worst-case error bars.

Because it is the sum $G_X^s(q^2)+G_X^{u,d}(q^2)$ from Eq.~(\ref{maindef}) that
is measured in
experiments, Fig.~\ref{fig:final} provides a theoretical error bar for the
extraction of $G_X^s(q^2)$.  As shown explicitly in Table III of
Ref.~\cite{KL},
modern experiments are already approaching this level of precision.
\vspace{-3mm}

\section*{Acknowledgements}

I am grateful to the organizers of PAVI06 for the opportunity to participate
in such an excellent conference, and to my isospin breaking collaborators,
Bastian Kubis and Nader Mobed.
The critical reading of this manuscript by
Bastian Kubis and Gerald Miller is greatly appreciated.
The work was supported in part by the Canada Research Chairs Program and
the Natural Sciences and Engineering Research Council of Canada.

\end{document}